\documentclass[twocolumn]{elsart5p}
\usepackage{graphicx}
\usepackage{amsmath}

\def\rmd{{\rm d}}
\def\om{\omega}
\def\prt{\partial}
\newcommand{\lsim}{\stackrel{\scriptstyle <}{\phantom{}_{\sim}}}
\newcommand{\gsim}{\stackrel{\scriptstyle >}{\phantom{}_{\sim}}}

\begin{document}

\begin{frontmatter}
\title{Fluctuations in non-ideal pion gas with dynamically fixed particle number}
\author[UMB]{E.E.~Kolomeitsev} \and
\author[MEPHI]{D.N.~Voskresensky}
\address[UMB]{Matej Bel University, Faculty of Natural Sciences,
     Tajovskeho 41, Banska Bystrica, SK-97401, Slovakia}
\address[MEPHI]{National Research Nuclear  University (MEPhI),Kashirskoe shosse 31, 115409 Moscow, Russia}

\begin{abstract}We consider a non-ideal hot pion gas with the dynamically fixed number of particles in the model with the $\lambda\phi^4$ interaction. The effective Lagrangian for the description of such a system is obtained after dropping the terms responsible for the change of the total particle number. Reactions $\pi^+\pi^-\leftrightarrow\pi^0\pi^0$, which determine the isospin balance of the medium, are permitted. Within the self-consistent Hartree approximation we compute the effective pion mass, thermodynamic characteristics of the system  and the variance of the particle number at temperatures above the critical point of the induced Bose-Einstein condensation when the pion chemical potential reaches the value of the effective pion mass. We analyze  conditions for the condensate formation in the process of thermalization of an initially non-equilibrium pion gas. The normalized variance of the particle number increases with a temperature decrease but  remains finite in the critical point of the Bose-Einstein condensation. This is due to the non-perturbative account of the interaction and is in contrast to the ideal-gas case. In the kinetic regime of the condensate formation the variance is shown to stay finite also.
\end{abstract}
\begin{keyword}
pion gas \sep Bose-Einstein condensation \sep fluctuations
\PACS
25.75.-q       
05.30.-d \sep  
24.60.Ky \sep  
24.10.Pa \sep  
\end{keyword}

\end{frontmatter}

\section{Introduction}

Multitude of pions is produced in central ultra-relativistic nucleus-nucleus collisions at SPS, RHIC and LHC energies~\cite{Afanasiev2002,Alt2008,Nayak:2012np,Abelev:2012wca,Adamczyk:2017iwn}.
Because of the kinematical separation of the light meson component or hadronization of quark-gluon plasma, a pion-enriched system can be formed at midrapidity. Spectra of produced pions are approximately exponential at intermediate transverse momenta, $m_\pi\lsim p_T\lsim 7\,m_\pi$, here and below $m_\pi=139$\,MeV is the pion mass, and $\hbar=c=1$. Therefore, one may assume that the pions stem from a hot fireball characterized by a temperature $T(\vec{r},t)$ and density $n (\vec{r},t)$, which expands until a thermal freeze-out.  The thermal freeze-out temperature is estimated to be $T_{\rm th} \sim 100-120$ MeV. Estimates~\cite{Goity:1989gs} show that at temperatures $T\lsim 130-140$\,MeV, the rate of pion absorption becomes smaller than the rate of pion rescattering. This implies that the total pion number can be considered as dynamically fixed during the subsequent fireball expansion before the thermal freeze out~\cite{Gerber:1990yb}. The time scale of chemical equilibration for $T\simeq 100-120$\,MeV, is estimated to be much longer, $\tau_{\rm ch}\sim 100$\,fm~\cite{Song1997,Pratt1999}, than that of thermal equilibration, $\tau_{\rm th}\sim $\,few fm, and of the time of the  fireball expansion, $\tau_{\rm exp}\sim 10-20$\,fm~\cite{Prakash1993}.

The measured $p_T$ spectra show an enhancement at low transverse momenta~\cite{Ableev}, which is attributed to a feed-down from resonance decays and a non-zero value of the pion chemical potential.  Already the first descriptions of pion spectra for collisions at $200A$GeV in the blast wave-model showed that the chemical potential of magnitude $\sim 120$--130\,MeV is required~\cite{Kataja:1990tp,mishust,mishust-2}. The recent fit~\cite{Begun:2014rsa} for collisions with $\sqrt{s_{NN}}=2.76$\,TeV obtained the chemical potential of pions $\mu\simeq  134.9$\,MeV that is just 80\,keV below the mass of the neutral pion.
Such a proximity of the pion chemical potential to the critical value of the Bose-Einstein condensation  in the ideal gas inspires speculations that the Bose-Einstein condensate (BEC) can be indeed formed in nucleus-nucleus collisions.
Although it was demonstrated in~\cite{GreinerGong} that the chemical potential cannot reach the critical value in the course of the quasi-equilibrium isentropic fireball expansion, non-equilibrium  overcooling effects~\cite{Voskresensky:1994uz,Voskresensky:1995tx,Voskresensky:1996ur} may provide a mechanism to drive pions to the BEC.
The condensation can also occur because of an additional injection of non-equilibrium pions from resonance decays~\cite{ornik}, decomposition of a ``blurred phase" of hot baryon-less matter \cite{Voskresensky:2004ux}, sudden hadronization of supercooled quark-gluon plasma~\cite{CC94} or a decay of the BEC of gluons preformed at the initial stage of heavy-ion collision, cf.~\cite{Blaizot:2011xf,Xu:2014ega}.

The presence of a phase transition (of the second or first order) implies an increase of fluctuations near the critical point,  as a general property manifested in various critical opalescence phenomena, cf.~\cite{LLP8}.
The proximity of the system created in nucleus-nucleus collisions to the Bose-Einstein condensation was argued in~\cite{Begun:2006gj,Begun:2008hq,Begun:2015ifa} to reveal itself through a divergence of the normalized variance. The results were derived for an ideal pion gas in the thermodynamical limit within both grand-canonical and micro-canonical ensembles.
Finite size effects  were also considered in~\cite{Begun:2008hq,Ayala:2016awt}. According to~\cite{Begun:2014rsa} the density of the ideal symmetric pion gas (with equal numbers of $\pi^+$, $\pi^-$ and $\pi^0$ mesons) at temperature $T_{\rm th}\simeq 138$\,MeV and chemical potential $\mu_{\rm th} \simeq
134.9$\,MeV is about $2.5n_0$ (here and below $n_0=0.16$~fm$^{-3}$ is the nuclear saturation density). Parameters obtained in~\cite{Kataja:1990tp} for O+Au collisions at 200$A$GeV,
$T_{\rm th}\simeq 167$\,MeV and $\mu_{\rm th} \simeq 134.9$\,MeV, give the  density for the ideal pion gas $\simeq 3.1\,n_0$. At such densities the pion gas is highly imperfect and an account of strong pion-pion interactions is required. On the experimental side some enhancement of the normalized variance was observed in pre-selected high pion-multiplicity events in $pp$ collisions at energies 50-70\,GeV~\cite{Kokoulina:2010zz,Kokoulina:2011ed}. However, a care should be taken when one compares theoretical estimations for thermal fluctuating characteristics with  actual measurements, which include experimental background fluctuations.

In~\cite{Voskresensky:1994uz}  properties of a pionic BEC were considered in the framework of the Lagrangian formalism with the $\lambda \varphi^4$
interaction. Utilizing  the Weinberg Lagrangian,~\cite{Kolomeitsev:1995we} studied the influence of the pion BEC for the one sort of pions on the kaon properties. The thermodynamical properties of the pion gas with the Weinberg Lagrangian were investigated in~\cite{BunatianWamb,Kolomeitsev:1996tv} within the self-consistent Hartree approximation. Reference~\cite{Kolomeitsev:1996tv} found the corresponding critical temperatures of the Bose-Einstein condensation.

This work is organized as follows. In next section we derive the effective Lagrangian describing the pion system with the dynamically fixed particle number. Then in section 3 we treat thermally equilibrated and not equilibrated pion systems in the self-consistent Hartree approximation. Macroscopic characteristics are considered in section 4.
In section 5 we  demonstrate that in the thermodynamical limit the normalized variance proves finite in the critical point of the pion Bose-Einstein condensation, if  the pion-pion interaction is taken into account self-consistently within the Hartree approximation. In section 6 we argue that the variance also remains finite during the non-equilibrium evolution of the system towards the BEC.

\section{Effective Lagrangian describing a system with dynamically fixed number of pions}

For the description of a pion system we use the Lagrangian density
\begin{align}\label{Wlag}
\mathcal{L} = \half(\,\prt\vec  \varphi\,)^2-\half m_{\pi}^2 {\vec \varphi}\,^2
-\quart\lambda
{\vec \varphi}\,^4,
\end{align}
where we assume a simplified $\lambda  \varphi^4$ model for the pion-pion interaction with constant $\lambda\sim 1$. Comparing with the leading terms of the Weinberg Lagrangian, we find $\lambda=m_\pi^2/2f_\pi^2\simeq 1.13$, where $f_\pi=93$\,MeV is the weak pion decay constant.
The isospin triplet $\vec \varphi=( \varphi_1, \varphi_2, \varphi_3)$ is associated with fields of the  positive $(\pi^+)$, negative $(\pi^-)$ and neutral $(\pi^0)$ pions:
$\pi^{\pm}=\frac{1}{\sqrt{2}}( \varphi_1\pm i \varphi_2)$, $\pi^0= \varphi_3$\,.
Positive and  negative pions, being introduced  as  particles and anti-particles, are described by the one complex field, $(\pi^-)^{\dagger}=\pi^+,$ where $(...)^{\dagger}$ means the Hermitian conjugation. Neutral pions, as self-conjugated particles, are represented by the real field $(\pi^0)^{\dagger}=\pi^0.$

\begin{figure}
\centering
\parbox{8.5cm}{\includegraphics[width=8.5cm]{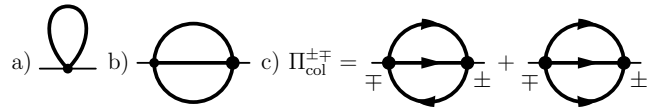}}
\caption{Contributions to the pion polarization operator: tadpole (a) and ``sandwich'' (b) diagrams. Thick lines represent the dressed pion propagators. The small blobs correspond to the bare vertex, whereas the big blobs stand for the dressed vertex. Diagram c) shows the pion self-energy being computed in the non-equilibrium diagram technique. The first diagram  corresponds to the pion elastic rescattering  and the second one, to the pion production and annihilation processes. Lines with $-$ and $+$ indices are the dressed non-equilibrium propagators  $iG^{-+}$ and $iG^{+-}$~\cite{Ivanov:1999tj,Voskresensky:1995tx}.}
\label{fig:polop}
\end{figure}

The interaction among the fields $\vec  \varphi$
modifies the pion properties  in medium. The resulting in-medium retarded pion propagator follows from the Dyson equation, $G(\om,k)=1/(G_0(\om,k)-\Pi(\om,k))$, where $G_0(\om,k)=1/(\om^2-k^2- m_{\pi}^2+i0)$ is the free   retarded pion propagator,
and $\om$ and $k$ stand for the  frequency and the momentum of the pion, and $\Pi$ is the full retarded pion  polarization operator determined by the two diagrams shown in Fig.~\ref{fig:polop}a,b. Note that in the ``tadpole graph'' (Fig.~\ref{fig:polop}a)
we have the bare coupling given by the Lagrangian (\ref{Wlag}), and the
vertex renormalization affects only one of the vertices in  the ``sandwich'' graph (Fig.~\ref{fig:polop}b).

Now we  apply the Lagrangian density (\ref{Wlag}) to the description of a pion system with  dynamically fixed number of particles. Particularly, the question is  how to define the number of neutral pions if the corresponding field $\pi^0$ in (\ref{Wlag}) is real and one cannot define a conserved 4-current for it. To proceed we transit to  new complex fields $\varphi_-$, $\varphi_+$, $\varphi_0$ corresponding to pionic modes with the positive frequencies
\begin{align}
\pi^-= \varphi_- +{ \varphi_+}^{\dagger},\,\,
\pi^+= \varphi_+ +{ \varphi_-}^{\dagger},\,\,\pi^0= \varphi_0 +{ \varphi_0}^{\dagger}.
\label{newfield}\end{align}
In a system with  fixed and, in general, different numbers of pions of each species the particle--anti-particle symmetry is lost, and we have now $ \varphi_-\neq \varphi_+^{\dagger}$, and $ \varphi_0$ is a complex field. In the second quantization the new fields of $\pi^-$, $\pi^+$, $\pi^0$ mesons possess the following operator representations in the quasi-particle approximation
\begin{align}
\hat \varphi_{i} &=\sum_{\vec{k}} {\hat a}_{\vec{k}}^{(i)}\, Z^{(i)}_{k}\, e^{-i\om_i(k) t+i\vec{k} \vec r}\,,\quad i=\pm,0\,.
\end{align}
We choose positive solutions of the dispersion relations, $\om_{\pm 0}(k)>0$, and, in general
case, $\om_-(k)\neq\om_+(k)\neq\om_0(k)$ with $Z^-_k\neq Z^+_k\neq Z^0_k$.
The operators $\hat a_{\vec{k}}^{(i)}$ define now the numbers of pions
of the corresponding $\sum_{\vec{k}}2\om_i Z_k^{(i)}
\overline{{\hat a}_{\vec{k}}^{(i)\dagger}{\hat a}_{\vec{k}}^{(i)}} = N_i \,,
$ where $\overline{(...)}$ denotes  averaging over the vector of state of the quantum
system with the Gibbs weight factor. The mean value $ \overline{{\hat a}_{\vec{k}}^{(i)\dagger}{\hat a}_{\vec{k}}^{(i)}}$ is determined by the
Bose distribution with the frequencies $\om_{i}$ and  chemical potentials $\mu_{i}$,
which fix the particle numbers $N_{i}$.

In terms of new fields (\ref{newfield}) the Lagrangian density (\ref{Wlag}) can be written as $\mathcal{L} = \mathcal{L}_{\rm fix}+\mathcal{L}^{'}$, where we separate terms conserving the number of pions of a certain sort
\begin{align}
\label{ourlag}
\mathcal{L}_{\rm fix}
=\mathcal{L}_{\pi^-}+\mathcal{L}_{\pi^+}+\mathcal{L}_{\pi^0}+\mathcal{L}_{\pi^+\pi^-\pi^0}
\,.
\end{align}
Here
$\mathcal{L}_{\pi^i} = |\prt \varphi_i|^2- m_{\pi}^2| \varphi_i|^2-\lambda| \varphi_i|^4\,$
is the Lagrangian density describing pions of a certain sort  $(i=\pm,0)$  and we neglect small differences in pion masses. Next term in~(\ref{ourlag}),
$\mathcal{L}_{\pi^+\pi^-\pi^0}= -4\lambda| \varphi_-|^2| \varphi_+|^2 -
2\lambda\left(| \varphi_+|^2+| \varphi_-|^2\right)| \varphi_0|^2,$
describes interactions of different pion species with each other via
reactions $\pi^+\pi^-\leftrightarrow\pi^+\pi^-$ and $\pi^{\pm}\pi^0\leftrightarrow\pi^{\pm}\pi^0$. These processes conserve the total
number of pions and also the number of pions of each sort.
The pion interactions contained in (\ref{ourlag}) give a contribution to the pionic polarization operator already in the first order in $\lambda$. Such a contribution is
depicted by the tadpole graph in Fig.~\ref{fig:polop}a.

The remaining part of the Lagrangian density is $ \mathcal{L}^{'}= \mathcal{L}_{2\leftrightarrow 2}+ \mathcal{L}_{3\leftrightarrow 1}\,, $
where $\mathcal{L}_{2\leftrightarrow
2}= -\lambda\big( \varphi_+ \varphi_- \varphi_0^{\dagger 2}
+ \varphi_0^2 \varphi_+^{\dagger} \varphi_-^{\dagger}\big)
$ corresponds to processes $\pi^0\pi^0\leftrightarrow\pi^+\pi^-$. They change  relative fractions of the pion species while keeping the total number of pions fixed. In thermal equilibrium these reactions imply  relations among  pion chemical potentials
\begin{align}
2\mu_0=\mu_++\mu_-\,.
\label{chempot}
\end{align}

In general, a pion system is characterized by two quantities: the total charge $Q$ and the total number of particles $N$. If say $Q>0$, then $Q$ pions out of  $N$ pions, are obviously  $\pi^+$s and other $N-Q$ pions form the electrically neutral sub-system with an arbitrary isospin composition. Reactions $\pi^0\pi^0\leftrightarrow\pi^+\pi^-$, having  occurred, bring
these $N-Q$ pions into the isospin equilibrium, in which every pionic species has the fraction $(N-Q)/3$. Taking into account the remained $Q$ charged pions we get the final isospin
composition of the  system
\begin{align}
N_{\pi^{+}}={\textstyle\frac13}(N+2Q),\quad
N_{\pi^{-}}=N_{\pi^0}={\textstyle\frac13}(N-Q).
\label{finalN}
\end{align}
After the replacement
$N_{\pi^+}\leftrightarrow N_{\pi^-}$ and $Q\to |Q|$ these expressions are valid also for $Q<0$.

The last term in $\mathcal{L}^{'}$, containing non-equal numbers of pion
creation and annihilation operators, $\mathcal{L}_{3\leftrightarrow 1}=\mathcal{L}( \varphi_i^{\dagger} \varphi_j \varphi_k \varphi_l; \varphi_i \varphi_j^{\dagger} \varphi_k^{\dagger} \varphi_l^{\dagger}), $
($i,j,k,l=\pm,0$)
corresponds to processes with a change of the number of pions, $\pi\leftrightarrow\pi\pi\pi$ symbolically. These processes bring the system to the chemical equilibrium.

The terms $\mathcal{L}_{2 \leftrightarrow 2}$ and $\mathcal{L}_{3\leftrightarrow 1}$
contribute to the pion polarization operator only in the second order in the coupling constant. They are represented by the sandwich diagram in Fig.~\ref{fig:polop}b with different possibilities for the pion species in  internal lines allowed by the charge conservation.

The sandwich diagrams written in terms of non-equilibrium Green's functions determine also the
collision integral in the pion kinetic equation~\cite{Voskresensky:1995tx,Ivanov:1999tj},
$I_{\rm col}=\int [\Pi^{-+}_{\rm col}G^{+-}-\Pi^{+-}_{\rm col}G^{-+}]\rmd\om/2\pi$, where
$G^{\pm\mp}$ are fully dressed non-equilibrium propagators, and $\Pi_{\rm col}^{\pm\mp}$ are determined by the non-equilibrium sandwich diagrams shown in Fig.~\ref{fig:polop}c, cf.~\cite{Voskresensky:1995tx,Ivanov:1999tj}.
The direction of internal $\pm$ lines in the diagrams ``c'' accounts for different processes: $\pi\pi \leftrightarrow \pi\pi$ if two lines got to the right and one to the left, see the first graph in Fig.~\ref{fig:polop}c, and $\pi \leftrightarrow \pi\pi\pi$ if all lines go to the right, see the second graph in Fig.~\ref{fig:polop}c. For $T\lsim m_\pi$ the real part of the retarded sandwich graph proves to be  small in   comparison to the tadpole contribution.  Therefore, we omit the former contribution. The imaginary part of the retarded sandwich diagram determines the rates of rescattering and absorption/production reactions. As we have discussed, the processes $\pi\pi\leftrightarrow \pi\pi$ responsible for the thermal equilibration occur essentially faster then processes $\pi \leftrightarrow \pi\pi\pi$ responsible for the chemical equilibration for $T\lsim m_\pi$. Therefore we also may drop the term $\mathcal{L}^{'}$ of the Lagrangian density and introduce instead non-vanishing pion chemical potentials fixing the pion numbers (\ref{finalN}) determined by reactions $\pi^+\pi^-
\leftrightarrow \pi^0\pi^0$. Doing so we assume that these reactions are operative on the time scale of the pion fireball expansion up to its breakup.

Thus we may now use the effective Lagrangian density (\ref{ourlag}) for the description of the pion gas of an arbitrary electric charge with the dynamically fixed number of particles.

\section{Self-consistent Hartree approximation}

\begin{figure*}
\centering
\includegraphics[height=4.5cm]{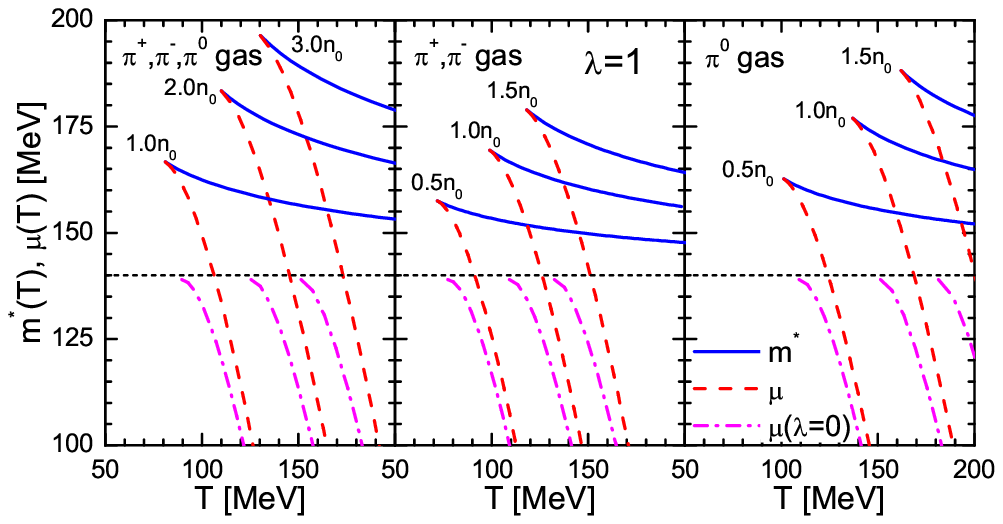}
\includegraphics[height=4.5cm]{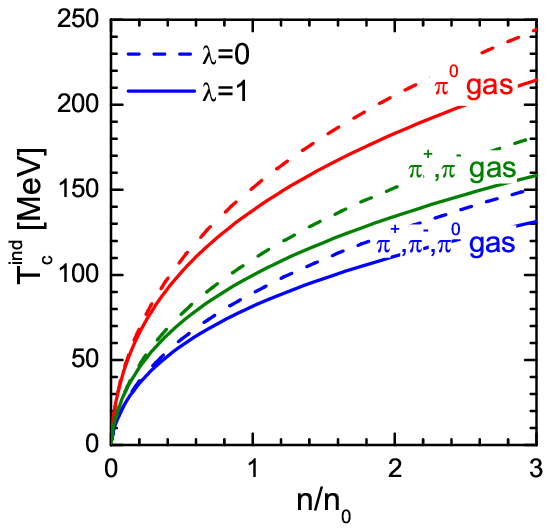}
\includegraphics[height=4.5cm]{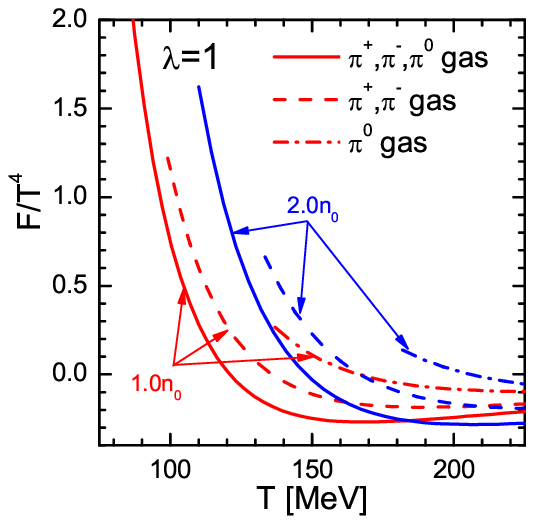}
\caption{{\it (Left panel)} Effective pion mass (solid lines) and chemical potential (dashed lines) as functions of temperature for the thermally equilibrated electrically neutral pion gas ($Q=0$) of different compositions ($i$)--($iii$): the symmetric $\pi^{\pm},\pi^0$ gas, the $\pi^{\pm}$ gas and the $\pi^0$ gas, for $\lambda =1$  and various densities.  Dotted line and dash-dotted lines show  the pion mass and the chemical potentials for the ideal gas. {\it (Middle panel)} Critical temperature of the induced Bose-Einstein condensation ($\mu(T_c^{\rm ind})=m^*(T_c^{\rm ind})$) for the pion gas of various compositions for $\lambda =0$ and $\lambda =1$.
{\it (Right panel)} Free energy density of the pion gas of various composition as a function of temperature for $\lambda=1$ and several densities.}
\label{fig:mmu-tc-f}
\end{figure*}

{\it Non-equilibrium system.}
Varying the action corresponding to the Lagrangian density (\ref{ourlag}) with respect to $\varphi_+^{\dagger}$, $\varphi_-^{\dagger}$, $\varphi_0^{\dagger}$ fields we obtain three coupled non-linear equations of motion. Direct solution of such a system of equations is a complicated numerical problem. Therefore here we solve this system within the self-consistent Hartree approximation, which is rather appropriate at excitation energies of our interest, $\bar{\epsilon}\sim T\lsim  m_{\pi}$. Within this approximation the behaviour of a certain particle is determined by an averaged interaction with surrounding particles.
Properties of particles in the environment are, in turn, determined by the same equation of motion as that for the considered pion.
Formally, we represent the field $\varphi_{\pm 0}$ as a superposition of some picked-out field $\tilde\varphi_{\pm 0}$ and an environmental field $\xi_{\pm 0}$, $\varphi_{\pm 0}\rightarrow\tilde\varphi_{\pm 0}+\xi_{\pm 0}$\,. Then we keep in  equations of motion only the terms   linear in  $\tilde\varphi_{\pm 0}$ and quadratic in $\xi_{\pm 0}$ (other terms vanish after the averaging). As the result we find equations of motion for the picked-out  fields in the Hartree approximation
\begin{align}
&\partial^2\tilde\varphi_{\pm} +m_\pi^2\, \tilde\varphi_{\pm}
+ \Pi_{\pm}\tilde\varphi_{\pm}
=0\,,
\nonumber\\
&\partial^2\tilde\varphi_0 +m_\pi^2\, \tilde\varphi_0\,+\,\Pi_{0}\tilde\varphi_{0}=0\,,
\label{EQM}
\end{align}
with the polarization operator $\Pi_i$:
\begin{align}
\Pi_+=\Pi_- &= 2\lambda (2\,\overline{|\xi_+|^2}+2\overline{|\xi_-|^2}+ \overline{|\xi_0|^2})\,,
\nonumber\\
\Pi_0 &=2\lambda (\overline{|\xi_+|^2}+\overline{|\xi_-|^2}+ 3\overline{|\xi_0|^2})\,,
\label{polop}
\end{align}
where
\begin{align}
\overline{|\xi_i|^2} =
\Big\langle\frac{f_i}{2\om_i(k)}\Big\rangle\,,\quad \langle...\rangle =\intop\frac{\rmd^3 k}{(2\pi)^3}\left(...\right)\,,
\label{d-def}
\end{align}
$f_i$ is a momentum distribution function, $i=\pm,0$.
The spectrum determined by (\ref{EQM}), (\ref{polop}) is
\begin{align}
\om_{i} = \sqrt{k^2+m_{i}^{*2}}\,\quad m_{i}^{*2} =m_\pi^2 + \Pi_{i}\,,
 \end{align}
with $m_{{0\pm}}$ as the pion effective mass. The partial densities of pions are determined as $n_i=\langle f_i \rangle\,.$

In general, the integrals (\ref{d-def}) contain  divergent contributions of quantum fluctuations owing to non-commutativity of creation-annihilation operators, which are to be renormalized by subtraction of the corresponding vacuum values. The remaining parts  prove to be rather small and we omit them.

Now to be specific consider the  electrically  neutral ($Q=0$) pion gas of three different compositions: $(i)$ the symmetric gas of all three species, $n_{\pm,0}=n/3$; $(ii)$ the gas of $\pi^+$s and $\pi^-$s, $n_{\pm}=n/2$, $n_0=0$; $(iii)$ the gas of $\pi^0$s, $n_0=n$, $n_{\pm}=0$. Then the  effective pion mass and density can be generally written as
\begin{align}
\label{efmasssym}
&m^{*2}=m_\pi^2+ g_m\lambda \Big\langle\frac{f}{2\om}\Big\rangle\,,
\\
& n = g_n \langle f\rangle\,\quad
\om = \sqrt{k^2+m^{*2}}\,.\nonumber
\end{align}
The coefficients $g_m$ and $g_n$ depend on the system we consider: $(i)$ $g_m=10$, $g_n=3$; $(ii)$ $g_m=8$\,, $g_n=2$\,; $(iii)$ $g_m=6$\,, $g_n=1$\,.

{\it System at thermal equilibrium.} For the thermally equilibrated system
\begin{align}
f_i=\frac{1}{e^{(\om_i(k)-\mu_i)/T}-1}\,,
\label{dens-int}
\end{align}
where the chemical potentials are interrelated by
\begin{align}
\mu_\pm=\mu\pm\mu_Q\,,\quad \mu_0=\mu
\label{murelat}
\end{align}
and are fixed by the total  and charge pion densities
\begin{align}
n = n_+ + n_- + n_0\,,\quad
n_Q \equiv n_+-n_- \,.
\label{dens}
\end{align}
In case of the thermally equilibrated electrically  neutral ($Q=0$) pion gas we get
$\mu_{\pm,0}=\mu$ in case  $(i)$,  $\mu_{\pm}=\mu$ in case $(ii)$, and   $\mu_{0}=\mu$
in case $(iii)$, and $\om_i=\om$ from (\ref{efmasssym}) in all three cases.

The dependence of the effective pion mass $m^*$ and the chemical potential on the temperature in the interacting electrically neutral pion gas ($Q=0$) is shown on the left panel in Fig.~\ref{fig:mmu-tc-f} for various gas compositions: cases ($i$)--($iii$) above. Because of the repulsive pion interaction the effective mass $m^*$ is larger than $m_\pi$ and increases with a temperature decrease. Consequently, the chemical potential also increases faster with decreasing temperature than in the free case ($\lambda=0$), grows over $m_\pi$  for temperature $T<T^*$ (for $T=T^*$ the chemical potential $\mu=m_\pi$) and equals to $m^*$ at the temperature $T_c^{\rm ind}$.
At this temperature an ``induced'' Bose-Einstein condensation occurs since at smaller temperatures the momentum distribution of pions in integrals (\ref{efmasssym})
would develop a pole. Following \cite{Kolomeitsev:1996tv} we should distinguish between the induced Bose-Einstein  condensation and a condensation occurring in a first-order phase transition at $T=T_c^{\rm (I)}$, $T_c^{\rm ind}<T_c^{\rm (I)}\leq T^*$, if the latter transition is energetically favorable. We postpone the study of such a possibility to a subsequent work. The critical temperature of the induced Bose-Einstein  condensation is shown on the middle panel in Fig.~\ref{fig:mmu-tc-f} as a function of the gas density for $\lambda=1$. We see that {\em the smaller is the number of components, the higher is the value $T_c^{\rm ind}$,} i.e. for cases ($i$)--($iii$) introduced above we have $T_c^{\rm ind}(iii)>T_c^{\rm ind}(ii)>T_c^{\rm ind}(i)$. In each of cases the critical temperature $T_c^{\rm ind}$ of the interacting pion gas is substantially smaller than that for the  ideal gas.

\section{Macroscopic characteristics}

A general expression for the energy-momentum tensor of a non-equilibrium system can be found in~\cite{Ivanov:1999tj}. For some types of interactions this quantity can be expressed fully  through the spectral functions of particles $A(\om,\vec{k})=-2\Im \Pi$, cf.~\cite{Voskresensky:2008ur}. Within the self-consistent Hartree approximation for spin-less bosons $A=2\pi\delta (\om^2 -m^{*2}-{k}^2)$.  As above we continue to consider systems with compositions given by cases ($i$)--($iii$) above. Within the self-consistent Hartree approximation the energy density $E$, the pressure $P$ and the entropy density are given by~\cite{Ivanov:1999tj,Voskresensky:2008ur}
\begin{align}\label{Eneq}
E&=g_n \Big\langle f\om +f\frac{m^2_\pi -m^{*2}}{4\om}\Big\rangle\,,
\nonumber\\
P&=g_n \Big\langle
f\frac{k^2}{3\om}-f\frac{m^2_\pi -m^{*2}}{4\om} \Big\rangle\,,\
\\
s&=g_n \Big\langle (1+f)\ln (1+f)-f\ln f\Big\rangle\,.\nonumber
\end{align}
In thermal equilibrium these quantities satisfy the consistency condition
\begin{align}
Ts&=E+P-\mu\, n\,,
\label{therm}
\end{align}
and the free-energy density is $F=E-Ts=\mu\,n-P\,.$

On the right panel in Fig.~(\ref{fig:mmu-tc-f}) we show the free-energy density as a function of the temperature for the interacting ($\lambda=1$) pion gas at thermal equilibrium with the different densities and compositions. We see that at fixed total density and for $Q=0$ the symmetric pion gas, case ($i$), has a lower free-energy than the gas of other isospin compositions, cases ($ii$) and ($iii$). {\em Thus  the system with $Q=0$ having initially an asymmetric composition will after a while come to the symmetric state with the equal number of pions of each species.}

Consider  initially non-equilibrium  system of pions with the already dynamically fixed particle number and the symmetric composition, case ($i$). We will assume that the  equilibration occurs much more rapidly  than the pion fireball expansion (as we mentioned in Introduction $\tau_{\rm th}\sim $few fm and $\tau_{\rm exp}\sim 10-20$~fm), so we assume the volume to be approximately fixed.
Let the initial energy density be $E_{\rm in}$ and the particle number density be $n_{\rm in}$. After a while (for ($t>\tau_{\rm th}$) the system reaches thermal equilibrium characterized by the pion  chemical potential  $\mu$ and the temperature  $T$. We assume that the energy, the particle number and the volume remain approximately conserved during the thermalization process. Then $E(\mu,T)=E_{\rm in}$ and $n(\mu,T)=n_{\rm in}.$
This system of equations is satisfied, if the chemical potential is smaller than the effective pion mass, $\mu<m^*(\mu,T)$, otherwise the system develops the BEC (provided the typical time for the appearance of the BEC is $\tau_{\rm BEC}<\tau_{\rm exp}$). The critical values of $E_{\rm in,c}$ and $n_{\rm in,c}$, for which the BEC appears, are connected by relations
$$E_{\rm in,c}=E(\mu_c(T),T)\,, \quad n_{\rm in,c}=n(\mu_c(T),T)\,,$$
where $\mu_c(T)$ is the solution of equation $\mu=m_\pi^*(\mu,T)$. They determine parametrically a line on the $(n_{\rm in},E_{\rm in})$-plane.

\begin{figure}\centering
\includegraphics[width=8.5cm]{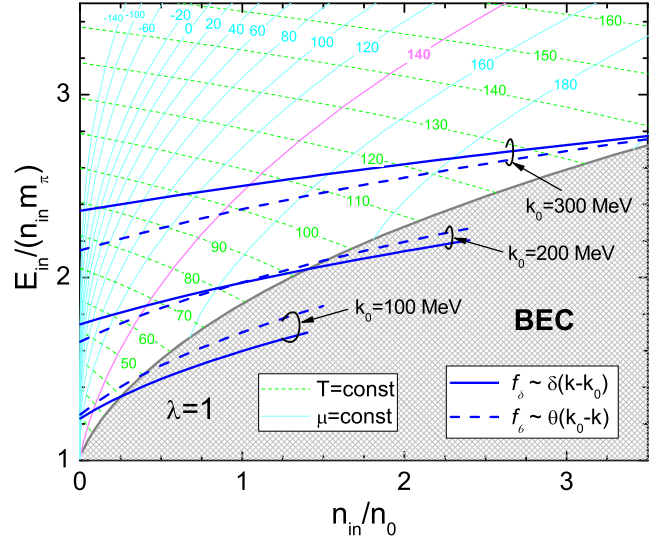}
\caption{Phase diagram of the interacting ($\lambda=1$) symmetric pion gas, case ($i$),  with an initial energy density $E_{\rm in}$ and particle number density $n_{\rm in}$. Hatched region is the region of parameters for which the system during its evolution reaches the BEC state. Thin dashed and solid lines correspond to the constant chemical potential and temperature (shown by labels in MeV units) after thermal equilibration of the system. Thick solid and dashed lines show the $E_{\rm in}$--$n_{\rm in}$ relation for the initial momentum distributions of non-equilibrium pions, $f_\delta$ and $f_\theta$, respectively. The results are presented for three different characteristic momenta of the distributions, $k_0=100$, 200, and 300\,MeV.
}
\label{fig:EinNin}
\end{figure}

Fig.~\ref{fig:EinNin} shows the phase diagram of the symmetric interacting pion gas, case ($i$), with an initial energy $E_{\rm in}$ and particle number $n_{\rm in}$ for $\lambda =1$ after its   thermal equilibration occurring at fixed volume, energy and particle number.
The critical line $T=T_c^{\rm ind}$ borders the hatched region of the Bose-Einstein condensation.
If the initial non-equilibrium pion distribution has the energy and particle densities within this region, then  the BEC is forming in the course of thermalization and continues to exist in the final thermal equilibrium state. If $E_{\rm in}$ and $n_{\rm in}$ lie outside the hatched region the system ends in a state with some $\mu$ and $T$, which are depicted by thin solid and dashed lines, respectively.

The initial momentum distribution of non-equilibrium pions depends on their source. We consider two simple choices: a $\delta$-function like distribution, $f_\delta(k)=\frac{2\pi^2\,n_{\rm in}}{3k_0^2}\delta(k-k_0)$, which may result from a ``blurred phase'' of hot baryon-less matter~\cite{Voskresensky:2004ux,Voskresensky:2008ur}, and a step $\theta$-function like  initial distribution in momenta  range $0<k<k_0$ used in  kinetic description of the BEC in~\cite{Semikoz:1994zp,Semikoz:1994zp-2}, $f_{\rm \theta}(k)=\frac{4\pi}{k_0^3}n_{\rm in}N_\Gamma\arctan[\exp(\Gamma(1-k^2/k_0^2))]$, with $\Gamma=5$ and $N_\Gamma\approx 0.963$.
Replacing $f\to f_\delta, f_{\rm \theta}$  in (\ref{efmasssym}) and (\ref{Eneq}) we obtain the line on the ($E_{\rm in}, n_{\rm in}$) plane for a given value of $k_0$.  These lines are shown in Fig.~\ref{fig:EinNin} (thick solid line for $f_{\delta}$ and thick dashed line for $f_{\rm \theta}$).
We see that the results for both distributions are qualitatively similar. For sufficiently small values of $k_0$ the critical densities are smaller than the estimated pion densities even at fireball freeze-out, e.g., $n_{\rm in,c} \sim 0.3\,n_0$ for $k_0=100$\,MeV and
$n_{\rm in,c} \sim 1.4\,n_0$ for $k_0=200$\,MeV. The BEC may appear for $n>n_{\rm in,c}$,
provided $\tau_{\rm BEC}<\tau_{\rm exp}$.

Reference~\cite{Semikoz:1994zp} estimated $\tau_{\rm BEC}$
studying the process of the Bose-Einstein condensation in gases at fixed particle number. The authors solved the non-relativistic quantum Boltzmann kinetic equation in the model with the $\lambda\varphi^4$ interaction. A similar problem for relativistic pions with dynamically fixed particle number in ultra-relativistic heavy-ion collisions was considered in~\cite{Voskresensky:1996ur}.  Following ~\cite{Semikoz:1994zp} the characteristic time scale for a change of the initial  distribution $f_{\rm \theta}(k<k_0)>1$
is $\tau_{\rm col} =\frac1{m^{*}}
\big(\frac{8k_0m^{*2}}{5 \sqrt{\pi} N_\Gamma\lambda n_{\rm in}}\big)^2$, and the typical time characterizing bosonization for low momenta is $t_{\rm bos}\sim 10\,\tau_{\rm col}$, and the BEC is forming for $t_{\rm BEC}\gsim 20\,\tau_{\rm col}$. Requiring  that $t_{\rm bos}\leq \tau_{\rm exp}\sim 20\,{\rm fm}$   we find that a strong bosonization for pions of small momenta occurs, if $n_{\rm in} \ge n_c \sim 2.5 \, n_0 (k_0/m_\pi) 
$\, for $\lambda=1$. For $k_0\lsim 0.7 m_\pi$ the BEC also starts to develop, if $n_{\rm in} \ge n_c \sim 2.5 \, n_0.$

\section{Particle number fluctuations in equilibrium}

In case of a uniform system of one-kind bosons in thermal equilibrium the normalized variance of the particle number $N$ is given by~\cite{LL1980}
\begin{align}\label{var-in}
{w}=
\frac{\overline{(\Delta N)^2}}{N}= \frac{T}{n}\frac{\partial n}{\partial \mu}\,.
\end{align}
The $\mu$- derivative is taken at fixed $T$.
The same relation is valid for the gas of various compositions, i.e. cases $(i\,\mbox{--}\,iii)$ considered above, when all pion species in the system have the same spectra. Then $N$ is the total pion number, and $n$ is the total pion density.
Taking the derivative with respect to $\mu$ we should take into account the implicit dependence of $m^*$ on it, cf.~(\ref{efmasssym}),
\begin{align}
{w}n
&=g_n\big\langle f( 1+f)\big\rangle -
g_n\Big\langle f\frac{1+f}{2\om}\Big\rangle\frac{\partial m^{*2}}{\partial \mu}\,,
\label{dndmu}\\
\frac{\partial m^{*2}}{\partial \mu} &=
g_m\lambda \frac{\partial}{\partial \mu}
\Big\langle \frac{f}{2\om}\Big\rangle\Big[1-g_m\lambda\frac{\partial}{\partial m^{*2}}
\Big\langle \frac{f}{2\om}\Big\rangle\Big]^{-1} \,.
\nonumber
\end{align}
With the help of the reduction formulas~(\ref{fluct-simpl}) and (\ref{part-deriv}) we obtain from Eq.~(\ref{dndmu})
\begin{align}
{w}(T)=
g_n \frac{T}{n}\Big( I_1 -\frac{g_m\frac{\lambda}{4}I_3^2}{1+g_m\frac{\lambda}{4}\, I_2}\Big)
\,,\label{var}
\end{align}
where
\begin{align}
I_1 =\Big\langle f\Big(\frac{1}{\om}+\frac{\om}{k^2}\Big)\Big\rangle
\,,
\,\,  I_2 = \Big\langle\frac{f}{\om \,k^2}\Big\rangle
\,,\,\,
I_3 = \Big\langle\frac{f}{k^2}\Big\rangle
\,.
\label{I123}
\end{align}
For $T\to T_c^{\rm ind}$, we have $\mu \to m^*$  (for $\lambda \neq 0$), and  the pion  momentum distribution  
behaves as $f\to \frac{2m^*T}{2m^*(m^*-\mu)+k^2}$ for $k\to 0$.
The values $I^{\rm (c)}_{1,2,3}=I_{1,2,3}(T\to T_c^{\rm ind})$ are
\begin{align}
I^{\rm (c)}_1=m^{*2}I^{\rm (c)}_2 = m^* I^{\rm (c)}_3
=\frac{T\, m^*}{2\sqrt{2} \pi  \sqrt{1- \mu/m^*}}\,.
\label{Iic}
\end{align}
Separating these divergent parts  we obtain  $\delta I_i=I_i-I_i^{\rm (c)}$, $i=1,2,3$, being finite for any $\mu\ge m^*$.

\begin{figure}
\centering
\includegraphics[height=4.5cm]{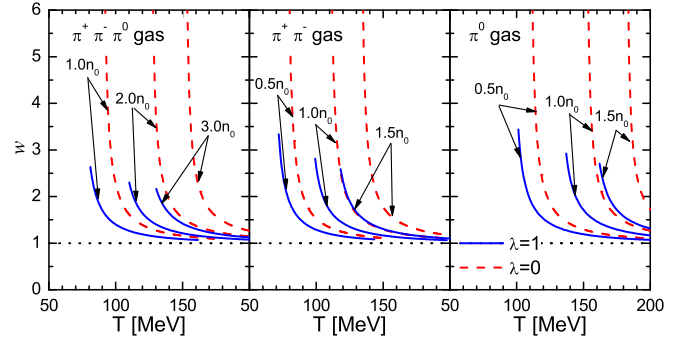}
\caption{Normalized variance of the pion number (\ref{var}) for the  charge-neutral ($Q=0$) interacting pion gas (for $\lambda=1$) and for the ideal pion gas ($\lambda =0$)  of various isospin compositions, cases ($i$)--($iii$), as a function of temperature for various densities.}
\label{fig:fluct}
\end{figure}

For the ideal pion gas ($\lambda =0$) we recover the standard expression $wn=g_n\,T\,I_1=g_n\,\langle f(1+f)\rangle$. Then at $T\to T_c^{\rm id}$ one has $\mu \to m_\pi$ and the variance $wn\to g_n\,T\,I_1^{\rm(c)}$ diverges as $wn\to
\frac{g_n\,T^2 m_\pi^{3/2}}{2^{3/2}\pi}(m_\pi-\mu)^{-\frac12}$ or in terms of the temperature ${w}\to \big(\frac{g_n\, T_c^{\rm (ind) 2}\, m_\pi}{2\pi n}\big)^2\frac{m_\pi}{T-T_c^{\rm (ind)}}$.

For the non-ideal gas the perturbation expansion in $\lambda$ is not valid for any $\lambda\neq 0$ since the integral $I_2$ in the denominator of~(\ref{var}) diverges. In the self-consistent Hartree approximation, separating convergent terms, we find
\begin{align}
{w}(T_c^{\rm ind})
&=\frac{g_nm^{*2}T_c^{\rm ind}}{g_m\lambda n}\,[4+g_m\lambda C(T_c^{\rm ind})] \,,
\label{varres}
\end{align}
where $C(T)=(\delta I_1 +m^{*2}\delta I_2 - 2\,m^*\delta I_3)/m^{*2}$.

In Fig. \ref{fig:fluct} we show the normalized variance of the number of pions for the interacting and ideal pion gases at $Q=0$ as functions of the temperature for several densities and various gas compositions. For the ideal gas the normalized variance diverges at the critical temperature of the Bose-Einstein condensation as was noted in~\cite{Begun:2008hq}. As argued in~\cite{Begun:2008hq}, the same divergence would be reflected in relative fluctuations of neutral vs. charged pions for high-multiplicity events.  Therefore it was suggested that a strong enhancement of the normalized variance can be considered as a clear signature of the BEC, although finite-volume effects smear the singularity. We see that {\em even for infinite volume the singularity disappears for the interacting gas provided the interaction is taken into account self-consistently.}

Speaking about particle-number fluctuations one has to be careful about the moment of the fireball evolution, at which fluctuations described here within the grand-canonical ensemble are formed: is it the chemical freeze-out $(n_{\rm ch},T_{\rm ch})$ or the thermal one $(n_{\rm th},T_{\rm th})$?
We believe that the answer depends on experimental conditions. Fluctuations of the total pion number, being measured within $4\pi$ geometry, reflect the state of the system at the chemical freeze-out since the total pion number does not change for $n<n_{\rm ch}$, $T<T_{\rm ch}$. On the other hand the grand-canonical formulation may continue to hold. First, fluctuations of the relative numbers of various pion species may be related to the thermal freeze-out since for $n\sim n_{\rm ch}$, $T\sim T_{\rm ch}$  the reactions $\pi^0\pi^0\leftrightarrow\pi^+\pi^-$  still occur.  Then, correlations of the volume and the density, being present  at thermal freeze-out, are  formally described by the same equation (\ref{var}).
The particle number fluctuations also  reflect the state of the pion fireball at the thermal freeze-out, if one measures correlations  between pions emitted from the fireball at different angles. However we remind that comparison of the results of idealized calculations with real measurements is very uncertain without a detailed study of experimental conditions.
Thus we may say that, only if indeed a significant growth of various pion-number fluctuation characteristics were observed, it could be associated with a closeness to the pion Bose-Einstein condensation either at the chemical freeze-out or at the thermal freeze-out, depending on the specifics of the measurement.

\section{Non-equilibrium effects on the particle number fluctuations}

We consider now if particle number fluctuations remain finite when the system enters a non-equilibrium regime towards the Bose-Einstein condensation, characterized by a strong dynamical enhancement of pion distribution function at small momenta. Since, as follows from the numerical analysis \cite{Semikoz:1994zp,Semikoz:1994zp-2}, the typical time characterizing elastic collisions for low momenta is by the order of magnitude larger than the  time $\tau_{\rm col}$, typical for collisions at averaged momenta $k\sim \sqrt{mT}$, for $t>\tau_{\rm col}$ we may consider the pion distribution at low momenta as still a non-equilibrium one, whereas for the higher momenta the initially overpopulated distribution is already the quasi-thermal one with $\mu (t)$ and $T(t)$ close to $m^*$ and $T^{\rm ind}_c$, respectively.  Then for $t\gg \tau_{\rm col}$ the fluctuation probability, $p$, can be determined by a minimal work needed to change  the chemical potential $\mu$ for fixed $n$ (for a constant volume), $P[\delta n]\propto \exp (- \frac{\partial \mu}{\partial n} (\delta n)^2/(nT))$. The normalized variance is then equal to ${w}^{\,\rm n.eq}\simeq \frac{T}{n} \frac{\partial n}{\partial \mu}$. So, we need to calculate $\frac{\partial n}{\partial \mu}$, now for the case of still non-equilibrium momentum distribution of pions at low momenta. For this we can proceed again from~(\ref{efmasssym}) with the replacement $f\to f_{\rm n.eq}$ in integrals. We assume that $f_{\rm n.eq}$ is a function of $\om-\mu$, since the chemical potential should enter through the gauge replacement $\om\to \om-\mu$. Thus we arrive first to (\ref{dndmu}), and then after the replacement $\frac{\partial f}{\partial \mu}\to -\frac{\partial f_{\rm n.eq}}{\partial \om}$  and the usage that
$\frac{\partial f_{\rm n.eq}}{\partial m^{*2}}=\frac{1}{2\om}\frac{\partial f_{\rm n.eq}}{\partial \om}$, we obtain  ${w}^{\rm n.eq}$. The latter quantity is given by the same relation  (\ref{var}) where one should  perform the replacements  $I_{1,2,3}\to \widetilde{I}_{1,2,3}$,
\begin{align}
\widetilde{I}_1 = \langle\psi\rangle\,,\quad
\widetilde{I}_2 = \Big\langle\frac{f_{\rm n.eq} + \om\psi}{\om^3}\Big\rangle\,,\quad
\widetilde{I}_3 = \Big\langle\frac{\psi}{\om} \Big\rangle,
\label{I-neq}
\end{align}
and $\psi=-\frac{\partial }{\partial \om} f_{\rm n.eq}$. Note also that the first term in the first Eq. ~(\ref{dndmu}), $g_n \langle f(1+f)\rangle$, coincides with the static structure factor calculated with the one boson loop for arbitrary non-equilibrium distributions. The second term is a more tricky, being obtained by re-summation of a sub-set of more complicated diagrams.

Numerical analysis~\cite{Semikoz:1994zp,Voskresensky:1996ur} shows that at low pion momenta, when  $|\om (k)-m^*|\ll m^*$, first a turbulence regime characterized by a constant particle flux into the region of small energies is formed and for $t >14\, \tau_{\rm col}$ the pion distribution acquires an self-similar form. For $t >19\, \tau_{\rm col}$ a singularity develops that signalizes the beginning of the formation of the BEC. At low momenta all these regimes are characterized by the power-low quasi-stationary distributions of the form
\begin{eqnarray}
f_{\rm n.eq} (t,k)\approx A(t)/k^{2\alpha}\approx
A(t)/(\om-m^*)^{\alpha}\,,\quad \alpha>1\,,
\label{turb}
\end{eqnarray}
where $A(t)$ is a slow function of time, being $k$-independent for $k\to 0$ and for  constant $\alpha$. For the thermal equilibrium $\alpha=1$, for the turbulent regime $\alpha=7/6$ and for the self-similar solution $\alpha=3/2$, cf.~\cite{Semikoz:1994zp,Voskresensky:1996ur}. Since the integrals in (\ref{I-neq}) are determined at small momenta we can use the distribution (\ref{turb}) for an estimate. With the distribution (\ref{turb})
we have $\psi \to 2m^*\alpha f_{\rm n.eq}/k^2$ and
the integrals $\widetilde{I}_i$ diverge at small momenta obeying the relations
$
\widetilde{I}_1=m^{*2}\, \widetilde{I}_2= m \widetilde{I}_3 \to
2\alpha m^*A(t)/(k_L^{2\alpha+1}(2\alpha+1))\,,
$ for $k_L\to 0$. 
Because of this universality relation among the integrals $\widetilde{I}_i$, which is the same as in (\ref{Iic}), the divergencies will cancel in ${w}^{\,\rm n.eq}$ leaving in the leading order ${w}^{\,\rm n.eq}=\frac{4 g_n m^{*2}}{g_m\lambda n\beta}+\dots$. {\em Thus, the pion self-interactions, taken self-consistently into account, keep the particle fluctuations finite also in the dynamics of the bosonisation and the BEC formation.}

\section{Conclusion}
In this work we considered the behaviour of the interacting hot pion gas with the dynamically fixed number of particles at temperatures above the critical point of the Bose-Einstein condensation. Such a system can be possibly formed in ultra-relativistic heavy-ion collisions. First, in the model of a simplified pion interaction ($\lambda\varphi^4$) we derived the effective Lagrangian, which describes a system with the dynamically fixed number of particles. The isospin balance is established by reactions $\pi^+\pi^-\leftrightarrow\pi^0\pi^0$, whereas pion production and absorption processes do not occur on the timescale of the fireball lifetime. The pion spectrum, thermodynamic characteristics and variance of the pion number are calculated within the self-consistent Hartree approximation. Conditions for the condensate formation in the process of thermalization of an initially non-equilibrium pion gas are analyzed.
As in the ideal gas case, in the interacting system at a fixed particle density (at fixed volume), the normalized variance of the particle number increases with a temperature decrease but remains finite in the critical point of the Bose-Einstein condensation.
The variance is also finite during the non-equilibrium evolution of the system towards the Bose-Einstein condensation.  We should also note that inelastic processes, which were ignored in our study, effects of a finite volume and of expansion of the pion system  may further smear the fluctuation effects due to enhancement of pion distributions at low momenta might be occurring in  heavy-ion collisions.

{\bf Acknowledgments.} We thank M.E.~Borisov for interest to this work and help.
The reported results were obtained within the state assignment by Ministry of Education and Science of the Russian Federation, project No~3.6062.2017/BY.
The study was supported by the Russian Foundation for Basic Research (RFBR) according to the research project No~16-02-00023-A, by Slovak grant VEGA-1/0469/15, by German-Slovak collaboration grant in framework of DAAD PPP project, and by THOR the COST Action CA15213.

\appendix
\section{Momentum integrals}\label{app}

For calculations of fluctuations we need integrals of the type $\big\langle f (1+f) \, g\big\rangle$, where $f$ is the momentum distribution function,  $g$ is another momentum-dependent function and angular brackets mean the momentum integration, see Eq. (\ref{d-def}). After integration by parts such expressions can be rewritten as
\begin{align}
\frac{1}{T}\big\langle f (1+f)\, g\big\rangle=
\Big\langle f \Big(\frac{g}{\om} + \frac{\rmd g}{\rmd \om}
+g \frac{\om}{k^2}\Big) \Big\rangle \,.
\label{fluct-simpl}
\end{align}
For thermal equilibrium when $f,g$ are functions of $T$, $\mu$, the integral $\big\langle fg\big\rangle$ is an explicit function of $T$, $\mu$ and $m^*$, which is finite for any $\mu>m^*$ if $gk^{2-\delta}\to const$ for  $k\to 0$ and $\delta>-1$, and $g\, e^{-k}\,k^\delta\to const$ for $k\to \infty$ and $\delta<-1$.
Partial derivatives of  $\big\langle fg\big\rangle$
with respect to  $T$, $\mu$ and $m^*$, keeping other two variables fixed, can be calculated with the help of relations
\begin{align}
\frac{\partial\langle f g\rangle}{\partial \mu}
&=\Big\langle f\Big(\frac{g}{\om} + \frac{\rmd g}{\rmd \om} +g \frac{\om}{k^2}\Big) \Big\rangle
\,,\quad
\frac{\partial\langle f g\rangle}{\partial m_\pi^{*2}}
=-\Big\langle\frac{f g }{2k^2}\Big\rangle\,,
\nonumber\\
\frac{\partial\langle fg\rangle}{\partial T}
&=\frac{1}{T}\Big\langle fg+
f(\om-\mu)\Big[\frac{g}{\om}+\frac{\rmd g}{\rmd \om} + \frac{g\om}{k^2}\Big]
\Big\rangle\,.
\label{part-deriv}
\end{align}

\end{document}